\pdfoutput=1 

\documentclass[11pt,letterpaper,twocolumn]{scrartcl}

\makeatletter
\DeclareOldFontCommand{\rm}{\normalfont\rmfamily}{\mathrm}
\DeclareOldFontCommand{\sf}{\normalfont\sffamily}{\mathsf}
\DeclareOldFontCommand{\tt}{\normalfont\ttfamily}{\mathtt}
\DeclareOldFontCommand{\bf}{\normalfont\bfseries}{\mathbf}
\DeclareOldFontCommand{\it}{\normalfont\itshape}{\mathit}
\DeclareOldFontCommand{\sl}{\normalfont\slshape}{\@nomath\sl}
\DeclareOldFontCommand{\sc}{\normalfont\scshape}{\@nomath\sc}
\makeatother

\usepackage[top=2.4cm, bottom=2.4cm, left=1.5cm, right=1.5cm,footskip=0.8cm]{geometry}

\usepackage[T1]{fontenc}
\usepackage[utf8]{inputenc}
\usepackage{abstract}

\usepackage[hidelinks]{hyperref}
\usepackage[parfill]{parskip}
\usepackage[separate-uncertainty = true,multi-part-units=single]{siunitx}
\usepackage{slashed}
\usepackage{microtype}
\usepackage{csquotes}
\usepackage{graphicx}
\usepackage{amsmath}
\usepackage{amssymb}
\usepackage{braket}
\usepackage{booktabs}
\usepackage{caption}
\usepackage{subcaption}
\usepackage{todonotes}
\usepackage{grffile}
\usepackage{enumitem}
\usepackage{afterpage}

\usepackage[numbers,sort&compress]{natbib}
\usepackage{multicol}

\usepackage{multirow}

\usepackage{fancyhdr}
\usepackage[yyyymmdd,hhmmss]{datetime}
\fancypagestyle{plain}
{
}
\usepackage{cleveref}

\usepackage[auth-sc,affil-it]{authblk}

\usepackage{scalefnt}
\newcommand{\abbrev}{\scalefont{.9}}

\newcommand{\NNNLL}{\text{\abbrev N$^3$LL}}
\newcommand{\NNNLLp}{\text{\abbrev N$^3$LL'}}
\newcommand{\NFLL}{\text{\abbrev N$^4$LL}}

\newcommand{\NNLO}{\text{\abbrev NNLO}}
\newcommand{\NNNLO}{\text{\abbrev N$^3$LO}}
\newcommand{\NLO}{\text{\abbrev NLO}}

\newcommand{\SM}{\text{\abbrev SM}}

\newcommand{\RG}{\text{\abbrev RG}}

\newcommand{\QCD}{\text{\abbrev QCD}}
\newcommand{\EW}{\text{\abbrev EW}}

\newcommand{\PDF}{\text{\abbrev PDF}}

\newcommand{\LHC}{\text{\abbrev LHC}}
\newcommand{\LHCb}{\text{\abbrev LHCb}}
\newcommand{\CMS}{\text{\abbrev CMS}}
\newcommand{\ATLAS}{\text{\abbrev ATLAS}}

\newcommand{\SCET}{\text{\abbrev SCET}}
\newcommand{\RGE}{\text{\abbrev RGE}}

\newcommand{\CuTeMCFM}{\texttt{CuTe-MCFM}}

\newcommand{\QED}{\text{\abbrev QED}}

\newcommand{\taucut}{\ensuremath{\tau_\text{cut}}}
\newcommand{\qtcut}{\ensuremath{q_T^\text{cut}}}

\newcounter{notecount}

\makeatletter
\renewcommand\maketitle{
	\begin{center}
		{\huge\bfseries\@title\par\vspace{0.3em}}
		{\scshape\@author, \@date}
	\end{center}
}
\makeatother

\fancypagestyle{firstpage}{%
	\lhead{}
	\rhead{FERMILAB-PUB-23-463-T}
}

\begin{document}
	
	\thispagestyle{firstpage}
	\title{\Large { Third order \QCD{} predictions for fiducial $W$-boson production}}

	\author[1]{John Campbell}
	\author[2]{Tobias Neumann}

	\affil[1]{Fermilab, PO Box 500, Batavia, Illinois 60510, USA}	
	\affil[2]{Department of Physics, Brookhaven National Laboratory, Upton, New York 11973, USA}
	
	\date{}
	\twocolumn[
	\maketitle
	
	\vspace{0.5cm}
	
	\begin{onecolabstract}
		\vspace{0.5cm} 
		
		Measurements of $W$-boson production at the \LHC{} have reached percent-level precision and 
		impose challenging demands on theoretical predictions. Such predictions directly limit 
		the precision of measurements of fundamental quantities such as the $W$-boson mass
		and the weak mixing angle.
		 A dominant source of uncertainty in predictions is 
		from higher-order \QCD{} effects.
		We present a calculation of $W$-boson production at the level of $\alpha_s^3$ at fixed 
		order and including 
		transverse-momentum resummation. 		
		 We further show predictions for a direct comparison with 
		low-pileup \ATLAS{} transverse-momentum and fiducial cross-section 
		measurements at $\sqrt{s}=\SI{5.02}{\TeV}$.
		We 
		discuss in detail the impact of modern \PDF{}s.
		 Our 
		calculation including the matching to $W$+jet production at \NNLO{} will be publicly 
		available the upcoming \CuTeMCFM{} release and allows for theory-data 
				comparison at the state-of-the-art level.

		\vspace{0.5cm}
	\end{onecolabstract}
	]
	
	\vspace{-5mm}
	
	\tableofcontents

	\section{Introduction}
	\label{sec:intro}
	
	The production of $W$ bosons at hadron colliders has one of the highest cross-sections of all
	Standard Model (\SM{}) processes.  Together with its relative ease of detection, through a large
	branching ratio into a lepton and neutrino, it has been measured with very high precision at
	multiple colliders. At the \LHC{}, measurements range from center of mass energies of 
	\SI{2.76}{\TeV} to \SI{13}{\TeV}, as performed by \ATLAS{} 
	\cite{ATLAS:2019fyu,ATLAS:2018pyl,ATLAS:2017irc,ATLAS:2016nqi}, \CMS{} 
	\cite{CMS:2016php,CMS:2014pkt,CMS:2011aa,CMS:2010svw} and \LHCb{} in the 
	forward region \cite{LHCb:2016zpq,LHCb:2016nhs,LHCb:2015mad,LHCb:2014liz,LHCb:2012lki}. Since 
	the first measurements 
	at the \LHC{},
	luminosity uncertainties have reduced from 2-3\% to 1\% \cite{CMS:2021xjt,ATLAS:2022hro}, 
	setting the upper bound on 
	the precision reached in current measurements \cite{ATLAS-CONF-2023-028}.	
	
	The precision of these analyses opens the door to the measurement
	of fundamental quantities appearing in the weak sector of the \SM{} Lagrangian,
	such as the $W$-boson mass 
	\cite{ATLAS-CONF-2023-004,ATLAS:2017rzl,LHCb:2021bjt,CDF:2022hxs,D0:2012kms,ALEPH:2013dgf}, the 
	weak mixing angle, as well as stringent 
	constraints on
	parton distribution functions (\PDF{}s) \cite{ATLAS:2021qnl,ATLAS:2016nqi,CMS:2013pzl} and 
	charge asymmetries \cite{CMS:2020cph,ATLAS:2019fgb,CMS:2016qqr,CMS:2012ivw,CMS:2011bet}.
	
However, the 
interpretation of 
these measurements --
and their ultimate precision -- depends crucially on the sophistication of the theoretical
predictions with which they are compared and analyzed.
	First \NNNLO{} predictions for $W$-boson production have been presented for total inclusive 
	cross-sections in ref.~\cite{Duhr:2020sdp}, where large cancellation effects between 
	initial-state channels have been observed that lead to significant \NNNLO{} 
	corrections of about $-2.5\%$.  	

	Fixed-order $\alpha_s^3$ $W$+jet predictions \cite{Gehrmann-DeRidder:2017mvr} ({\abbrev 
	NNLOjet}) have been matched to the 
	{\abbrev RadISH} resummation \cite{Bizon:2017rah,Monni:2016ktx} in 
	ref.~\cite{Bizon:2019zgf} at the level of \NNNLL{} and compared to Pythia results. Higher-order 
	transverse-momentum resummation at the level of \NNNLLp{} matched to lower-order $\alpha_s^2$ 
	predictions has also been studied in refs.~\cite{Ju:2021lah,Camarda:2023dqn}. Recent studies of 
	threshold resummation in rapidity distributions were presented in 
	refs.~\cite{Das:2023bfi,Ajjath:2020rci}.
	Fixed-order \NNNLO{} predictions for $W$-boson production have been presented in 
	ref.~\cite{Chen:2022lwc} (transverse mass distribution, rapidity, charge asymmetry), which
	also includes a Tevatron study using fiducial cuts.
	
	Generally the residual \QCD{} truncation uncertainties at the level of $\alpha_s^3$ are 
	estimated to be at the level of $1-2\%$.
	Apart from \QCD{} effects, other Standard Model effects play a role at the level of 1\% 
	precision. Among these, mixed \QCD{}x\EW{} corrections were reported in 
	ref.~\cite{Buonocore:2021rxx,Behring:2020cqi,Dittmaier:2020vra,Dittmaier:2014qza,Dittmaier:2015rxo}
	and \QED{}-\QCD{} transverse-momentum resummation in ref.~\cite{Autieri:2023xme}. 
	
	While all of these recent higher-order corrections are important for percent-level comparison 
	with kinematic measurements, they are crucial to improve the $W$-boson mass 
	measurement. 
	In particular,	even though few measurements have been performed of the $W$-boson
	transverse momentum directly \cite{ATLAS-CONF-2023-028,CMS:2016mwa,ATLAS:2011fpo},
	 it is a key part in the $W$-boson mass analyses.
	A comprehensive review of how theoretical contributions and uncertainties impact 
	the $W$-boson mass measurement was presented in ref.~\cite{CarloniCalame:2016ouw} (2016). An 
	estimate for the impact of mixed \QCD{}x\EW{} corrections has since 
	also been performed~\cite{Behring:2021adr}.

	In this paper we present a calculation of fully differential $W$-boson production at \NNNLO{} 
	fixed-order and including transverse-momentum resummation up to a logarithmic order of 
	\NFLL{} matched to 
	\NNNLO{} fixed-order.\footnote{The logarithmic accuracy of \NFLL{} is only reached within the limitations of 
	current \NNNLO{} \PDF{} approximations.}
	Together with our calculation of $Z$ production at $\alpha_s^3$ \cite{Neumann:2022lft}, it 
	completes the set of two crucial processes entering many Standard Model precision 
	analyses, for example the $W$-boson mass determination. It allows for \QCD{} predictions at the 
	highest level with an independent implementation of higher-order ingredients and resummation. 
	The availability of multiple calculations in different resummation formalisms enables reliable 
	uncertainty estimates.
        In this paper we  compute predictions for comparison with the 
	$\sqrt{s}=\SI{5.02}{\TeV}$ 
	\ATLAS{} analysis from ref.~\cite{ATLAS-CONF-2023-028}.
	The calculation will be made publicly available, so that its results can be used for
	future analyses, in the upcoming release of \CuTeMCFM{}.	
	
	In \cref{sec:setup} we describe our setup and cross-checks performed on our calculation. In 
	\cref{sec:fidcross} we first discuss predictions of the fiducial cross-section and compare with 
	the measurement. In particular we focus on the impact of \PDF{}s. We move on to differential 
	distribution in \cref{sec:results}, where we discuss the $W$-boson transverse-momentum 
	distribution (\cref{sec:ptW}) and transverse mass and charged lepton transverse-momentum 
	distributions (\cref{sec:mtptl}). Since the impact of \PDF{}s is significant, we discuss their 
	impact on these distributions in \cref{sec:diffpdf}. We conclude and present an outlook in 
	\cref{sec:conclusion}.
	
\section{Setup}
\label{sec:setup}

We implement \QCD{} corrections to $pp \to W(\to e \nu)$ production at fixed-order and including 
the effect of $q_T$-resummation in \CuTeMCFM{} 
\cite{Becher:2020ugp,Campbell:2019dru,Neumann:2021zkb,Neumann:2022lft}. We 
achieve $\alpha_s^3$ fixed-order and transverse momentum 
renormalization-group-improved (\RG{}-improved) logarithmic accuracy, counting 
$\log(q_T^2/Q^2)\sim1/\alpha_s$. Note that the logarithmic accuracy of \NFLL{} ($\alpha_s^3$) 
relies on the 
availability of \NNNLO{} \PDF{}s, in particular on the four-loop {\abbrev DGLAP} evolution. \PDF{}s 
at this order are so far only available by the {\abbrev MSHT} group in an 
approximation \cite{McGowan:2022nag}. The ingredients of our calculation and the checks we have 
performed are detailed below.
	
\paragraph{Resummation.}
	
The implementation of the resummation formalism follows our study on $Z$ production 
\cite{Neumann:2022lft}, it is based on the \SCET{}-derived $q_T$-factorization theorem developed in 
refs.~\cite{Becher:2010tm,Becher:2011xn,Becher:2012yn} and originally implemented to \NNNLL{} as 
\CuTeMCFM{} in ref.~\cite{Becher:2020ugp}. Large logarithms $\log(q_T^2/Q^2)$ are 
resummed through \RG{} evolution of hard- and beam functions in a small-$q_T$ factorization 
theorem. Rapidity logarithms are directly exponentiated through the collinear-anomaly 
formalism.

The transition from the resummation which is valid only at small $q_T$, to fixed-order predictions 
at large $q_T$, is achieved through the use of a transition function that smoothly interpolates between those two 
regions \cite{Becher:2020ugp}. The overlap between fixed-order and resummation is removed through a 
fixed-order expansion of the factorization formula. The difference between these two parts is 
referred to as matching corrections. While for $Z$ production matching corrections quickly approach zero 
for $q_T\to0$,
even at the level of $\alpha_s^3$, we find that for $W$ production they are at the level of 
a few percent even at relatively small $q_T$ (as described in detail in \cref{sec:cutoffeffects}
below).

The higher-order ingredients in the resummation calculation are identical to those in $Z$ production. We 
briefly summarize the most important ones here, and refer for more details to our implementation of 
$Z$ production \cite{Neumann:2022lft}.
Three loop transverse momentum dependent beam functions that allow for resummation at the level of 
\NNNLLp{} have been calculated in refs.~
\cite{Luo:2020epw,Ebert:2020yqt,Luo:2019szz}. We include the four loop rapidity anomalous dimension 
\cite{Duhr:2022yyp,Moult:2022xzt}, which together with \NNNLO{} \PDF{}s allows for \NFLL{} 
resummation. The five-loop cusp anomalous dimension is numerically negligible, which is true 
already at the four-loop level.

The formalism of refs.~\cite{Becher:2010tm,Becher:2011xn,Becher:2012yn} further employs an 
additional 
power counting that improves the resummation at small $q_T$ 
\cite{Becher:2011xn} and avoids a Landau pole prescription, as the relevant scales are always set 
in the perturbative regime. Implementing this requires the inclusion of higher-order terms of the 
beam functions which were reconstructed from the beam-function \RGE{}s in 
ref.~\cite{Neumann:2022lft}. 

The hard function entering the factorization formula consists of {\abbrev 
$\overline{\text{MS}}$}-renormalized virtual corrections. For $Z$ production these are more 
complicated due to separate singlet and axial-singlet contributions, which are not present for $W$ 
production. The hard function is therefore solely given by the three-loop (vector) form-factor 
\cite{Gehrmann:2010ue,Baikov:2009bg,Lee:2010cga}.

\paragraph{Fixed order.}

To obtain fixed-order \NNNLO{} results we use $q_T$ slicing, which was implemented in 
ref.~\cite{Neumann:2022lft} using the same factorization theorem and ingredients as laid out above 
for the resummation.

The $W$+jet \NNLO{} calculation, which is necessary above the $q_T$ slicing cutoff, is based on 
ref.~\cite{Boughezal:2015dva} using 1-jettiness 
subtractions \cite{Gaunt:2015pea,Boughezal:2015dva,Stewart:2010tn} and the 1-jettiness 
soft-function of refs.~\cite{Campbell:2017hsw,Boughezal:2015eha}.
We have thoroughly cross-checked all elements of this calculation. For example, we find agreement 
between all amplitude expressions and Recola \cite{Denner:2017wsf}. Further checks were performed 
as part of the validation of $Z$+jet production, for example a re-implementation of the non-singlet 
hard function using refs.~\cite{Gehrmann:2011ab,Garland:2002ak,Gehrmann:2002zr} that was 
originally taken from the code {\abbrev PeTeR} \cite{Becher:2011fc,Becher:2013vva}.
We have identified and corrected small inconsistencies in the original implementation of \NLO{} 
subtraction terms in the $W+2$~jet process~\cite{Campbell:2002tg}, a component in our above-cut 
calculation. We have further checked our final \NNLO{} $W$+jet results against 
a fully independent calculation presented in ref.~\cite{Gehrmann-DeRidder:2018kng}.

\subsection{Cutoff effects.}
\label{sec:cutoffeffects}

We study fixed-order and resummed results, which are both affected by cutoff effects in different 
ways. The resummed calculation requires a cutoff of the matching corrections, while our \NNNLO{} 
fixed-order 
calculation is based on a nested slicing approach, regularizing \NNNLO{} singularities using $q_T$ 
subtractions. In both cases the \NNLO{} $W$+jet calculation is performed using 1-jettiness slicing, 
and therefore, unlike for the local antenna subtractions used in the {\abbrev NNLOjet} calculation 
\cite{Gehrmann-DeRidder:2017mvr}, we have to particularly pay attention to residual slicing
cutoff effects.

For the plots here and throughout the results section, we use the cuts of a recent \ATLAS{} 
study \cite{ATLAS-CONF-2023-028} shown in eq.~\ref{eq:fidcuts} in \cref{eq:fidcuts}. The choice of 
symmetric $q_T$ cuts on the $W$ decay leptons makes the calculation numerically challenging 
\cite{Chen:2022cgv}, even when including linear power corrections in the $q_T$ slicing method.

Unlike for $Z$-boson production \cite{Neumann:2022lft} the cutoff effects are not negligible at the 
order of $\alpha_s^3$ for cutoff values of \SIrange{3}{5}{\GeV} that we achieve here. We therefore 
take care to display the limitations of this throughout our results.

\paragraph{Cutoff effects in the resummation.}

\begin{figure}
	\centering
	\includegraphics[width=\columnwidth]{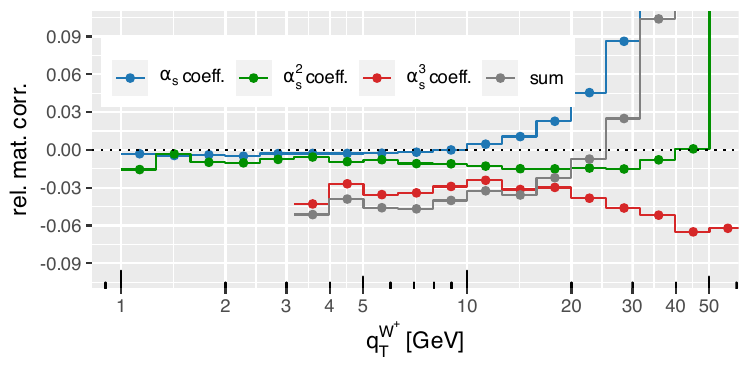}
	\caption{Matching corrections for $W^+$  production at $\sqrt{s}=\SI{5.02}{\TeV}$ with fiducial 
	cuts as in \cref{eq:fidcuts}.}
	\label{fig:matchcorrfid_main}
\end{figure}

We first discuss cutoff effects in the matched resummed result. In \cref{fig:matchcorrfid_main} we 
show the matching corrections relative to the purely resummed 
$\alpha_s^3$ result at different orders in $\alpha_s$. While they are small and quickly approach 
negligible levels at lower orders, the $\alpha_s^3$ corrections are substantial.

At lower orders we use a matching corrections cutoff of \SI{1}{\GeV}, with negligible impact on the 
results, while at $\alpha_s^3$ we use a cutoff of \SI{3.16}{\GeV} and a $q_T$-dependent 
dynamic jettiness cutoff of the \NNLO{} $W$+jet calculation of $\frac{0.03}{1050}\sqrt{q_T^2 + 
	m_{ll}^2}$, so about \SI{0.002}{\GeV} at small $q_T$. Lower values of $q_T$ would require 
	smaller values of a jettiness cutoff, significantly increasing computational resources.

From  \cref{fig:matchcorrfid_main}, the matching corrections are still about 3\% 
around our cutoff of \SI{3.16}{\GeV}.
We estimate the uncertainty due to missing matching corrections by multiplying 
the purely resummed result integrated up to \SI{3.16}{\GeV} by three percent. The impact of this 
is different in various kinematical distributions and also depends on the binning. For example in 
the $W$-boson $q_T$ distribution in \ATLAS{} binning \cite{ATLAS-CONF-2023-028} the first bin 
ranges from \SIrange{0}{7}{\GeV}. The 
effect of neglected matching corrections in this bin is up to about 1.5\%, while there is no such 
error in the other bins. Overall, the effect of neglecting matching corrections is therefore not 
negligible and needs to be included in uncertainty estimates.

In distributions other than the $W$-boson $q_T$ this effect is smeared 
out and we include it as an additional error bar. Since we know that the effect is likely to lead to 
a reduction of the cross-section, we display the error bar only in the downwards direction.

The size of the matching corrections also indicates where a transition function needs to switch 
between the resummed and fixed-order calculations. The matching corrections become sizeable beyond 
\SI{40}{\GeV}, which motivates a transition function as detailed in 
ref.~\cite{Becher:2020ugp} using $x_T^{\text max} = (q_T^{\text max}/M_Z)^2$
with $q_T^{\text max}$ in the range $35$ to \SI{60}{\GeV}. With this choice we find that transition 
uncertainties, obtained by varying the transition function ($x_T^\text{max} = 0.2,0.4,0.6$), are 
then comparable to uncertainties in the fixed-order and resummation regions. They are therefore 
insensitive to the precise range and shape of the transition.

\paragraph{Cutoff effects at fixed order.}

Our \NNLO{} and \NNNLO{} fiducial fixed-order cross-sections are computed using $q_T$ slicing.
For the resummed calculation linear power corrections are included automatically through a recoil 
prescription \cite{Ebert:2020dfc,Becher:2019bnm}.  In the fixed-order case they have to be added 
separately, although this is straightforward \cite{Buonocore:2021tke}.

\begin{figure}
	\centering
	\includegraphics[]{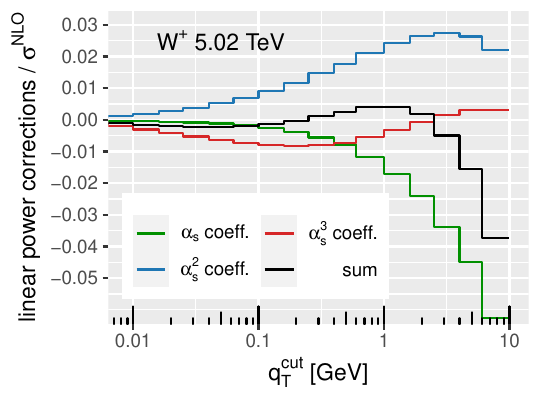}
	\caption{Linear power corrections for $W^\pm$ production at $\sqrt{s}=\SI{5.02}{\TeV}$ 
		relative to the \NLO{} cross-section, with input parameters as in \cref{sec:results}.}
	\label{fig:linpc}
\end{figure}

The size of the power corrections for the $\alpha_s^k$ coefficient relative to the full 
$\alpha_s$ result is shown in \cref{fig:linpc} for $W^+$production, as a function
of the $q_T$ slicing cutoff,  $\qtcut{}$.\footnote{The relative power corrections for $W^-$ 
production
are virtually identical.}
At \NNLO{} we 
use a $q_T$ slicing cutoff of $\SI{1}{\GeV}$
with linear power corrections on the cross-section at the level of 2\%. For the \NNNLO{} 
coefficient we use a cutoff of $\SI{3.16}{\GeV}$ and the power corrections are a few per-mille.
 The large size 
of the power corrections is an effect of the symmetric lepton cuts \cite{Chen:2022cgv}.

\begin{figure}
	\centering
	\includegraphics[width=\columnwidth]{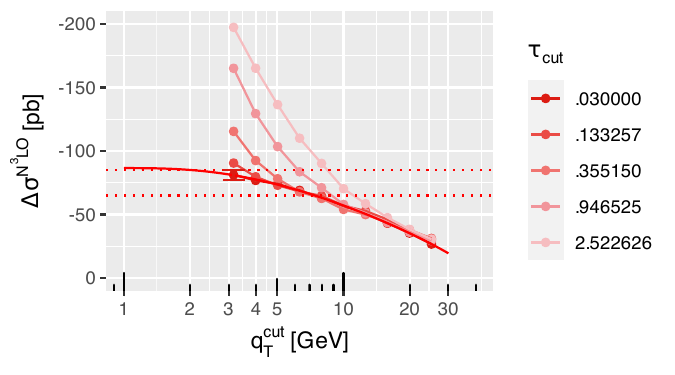}
	\caption{$q_T^\text{cut}$ and $\taucut$ extrapolation of the fiducial \NNNLO{} 
		$W^+$ cross-section coefficient at \SI{5.02}{\TeV}
		(not 
		including the small linear power corrections). The error bar denotes the estimated 
		numerical uncertainty. The solid red line is the result of a fit. The dotted red lines 
		are an estimate of our overall uncertainty. }
	\label{fig:taucutfid_main}
\end{figure}

Our final fiducial \NNNLO{} corrections are obtained from a $\qtcut{}$ extrapolation, taking into 
account that subsequently smaller $\taucut{}$ values are necessary for small $\qtcut{}$ in the 
inner $W$+jet \NNLO{} calculation. This is shown in \cref{fig:taucutfid_main} for $W^+$ production.
The dependence for $W^-$ is qualitatively the same. Note that the $\taucut$ dependence is modified 
by the dynamic choice, our default is $\tau \propto \sqrt{q_T^2 + m_{ll}^2}$. 
Other choices for the functional dependence of $\tau$, such
as $\tau \propto q_T$, may lead to improved performance.  We leave a detailed investigation of
such choices to a future publication.

The solid red line is 
from one possible fit of the expected asymptotic behavior. Smaller values of $\qtcut{}$ would be 
desirable, 
and there is some uncertainty of the fit that can be exposed for example by varying the number of 
terms that are included. The overall pattern is that for too large $\taucut$ the cross-section diverges 
towards more negative values. We find that at the smallest $\qtcut{}$ value the two smallest 
$\taucut{}$ values are not fully overlapping and therefore we might miss a further slight 
flattening of the curve that would impact the fit. In addition to the fit uncertainty,
the red error bar denotes our 
numerical uncertainty that affects all points.

Overall we assign a $0.5\%$ technical uncertainty on the fiducial 
cross-section, which is 
visualized by the dotted red bars in \cref{fig:taucutfid_main}. Our central value is obtained at 
$\qtcut=\SI{4}{\GeV}$. This technical uncertainty includes our numerical 
statistical uncertainty and the uncertainty in $\qtcut{}$ and $\taucut{}$ extrapolation. It 
is large enough to cover the data points from $\SI{3.16}{\GeV}<\qtcut{}<\SI{7}{\GeV}$.

As an additional check of our setup we worked to reproduce the total inclusive 
cross-section that can be calculated conveniently with the code \texttt{n3loxs} 
\cite{Duhr:2020sdp,Baglio:2022wzu}. We find agreement, but unfortunately this calculation is 
numerically particularly challenging in our nested slicing approach. In the fiducial case the power 
corrections 
are more manageable, which allowed us to obtain results with a numerical and slicing error of less 
than $0.5\%$. In the total inclusive case the power corrections are larger, but also the numerical 
integration turns out to be more difficult. We find agreement within a combined uncertainty of 
about 
1\%. Given that the total \NNNLO{} 
corrections are about $-1.7\%$ for the choice of invariant mass range and input parameters that we 
took, this is 
not as strong of a cross-check as we would have liked.
We also computed the total inclusive cross-section including the effect of $q_T$ resummation and 
find an inconclusive difference of $-1.8\%\pm 1.2\%$ (here also an uncertainty from the transition 
function enters).

We used about 50000 NERSC Perlmutter core hours (64 nodes for 12 hours) 
to compute the total inclusive \NNNLO{} 
cross-section to the precision of 1\%. While these are not huge resources in the context 
of current 
calculations \cite{FebresCordero:2022psq} which can go into the millions
of core hours, reducing the numerical error 
by a factor of one third 
would already ten-fold these resource requirements, since Monte-Carlo integration uncertainties 
decrease like the square root asymptotically. The 
large power corrections further demand smaller cutoffs to reduce the slicing uncertainty, which also
contributes sizeably in the total.
Accounting for power corrections in the $W+\text{jet}$ \NNLO{} calculation provides a possible
path for reducing the overall slicing uncertainty \cite{Boughezal:2019ggi,Caola:2022ayt}.
However, ultimately a more efficient approach such as a local subtraction procedure,
and ideally local \NNNLO{} subtractions, will be necessary to substantially reduce these 
uncertainties.

\section{Results}
\label{sec:results}

In this section we show fiducial cross-sections and distributions for $W^+$ production at 
$\sqrt{s}=\SI{5.02}{\TeV}$ corresponding to the \ATLAS{} analysis in 
ref.~\cite{ATLAS-CONF-2023-028}. The fiducial cuts for this analysis are,
\begin{eqnarray}
&& q_T^\ell > \SI{25}{GeV} \,, \qquad q_T^\nu > \SI{25}{GeV} \,, \qquad
 \left| \eta^\ell \right| < 2.5\,, \nonumber \\
&& m_T = \sqrt{2 q_T^\ell q_T^\nu (1-\cos\Delta\phi_{\ell\nu})} > \SI{50}{GeV}\,.
\label{eq:fidcuts}
\end{eqnarray}
Note that the theory predictions used in that analysis are at a 
lower order and do not include any uncertainties.
We include the corresponding plots for $W^-$ 
production in \cref{app:additional} since overall the relative perturbative corrections 
are very similar. 

For all our predictions we use the $G_\mu$ scheme with $m_W=\SI{80.385}{\GeV}$, 
$m_Z=\SI{91.1876}{\GeV}$ and $G_F=\SI{1.6639e5}{\GeV^{-2}}$, as well as 
$\Gamma_W=\SI{2.091}{\GeV}$. Different scheme choices can be used to estimate the effect of 
higher-order electroweak corrections \cite{Alioli:2016fum} which we do not consider here.
We examine the impact of various \PDF{} sets, extracted at different perturbative orders,
in our study and will make clear which 
\PDF{} set is used in each prediction.

	
In the following, the label $\alpha_s^k$ at fixed-order denotes {\abbrev N$^k$LO}, while for 
the 
resummed cross-section it denotes {\abbrev N$^{k+1}$LL+N$^k$LO}, that is $\alpha_s^k$ in an 
\RG{}-improved power counting where large logarithms $\log(q_T^2/Q^2)$ are counted as $1/\alpha_s$. 
Note 
that only the prediction 
using the \texttt{MSHT20an3lo} \PDF{} set  \cite{McGowan:2022nag} reaches approximately \NFLL{} 
logarithmic accuracy,
within the limitations of its approximations. For the \NNLO{} \PDF{} sets the effect from 
missing four-loop splitting functions degrades the formal accuracy to \NNNLLp{}, see 
ref.~\cite{Neumann:2022lft}.

As well as the fiducial cross section, we choose to discuss the $W$-boson transverse momentum,
charged-lepton 
transverse momentum and $W$-boson transverse mass distributions, which are of particular interest 
for the $W$-boson mass analyses.  
	
	\subsection{Fiducial cross-sections}
	\label{sec:fidcross}
	
	We start with a discussion of the total fiducial cross-section that we compare with the 
	recent \SI{5.02}{\TeV} \ATLAS{} measurement \cite{ATLAS-CONF-2023-028}. We compare 
	predictions of different perturbative orders in $\alpha_s$ at fixed order 
	as well as including the effect of $q_T$ resummation at the respective 
	logarithmic order.
        Even at the level of the fiducial cross section this is interesting because,
	in the presence of symmetric lepton $q_T$ cuts such as those in this analysis
	(c.f. equation~(\ref{eq:fidcuts})), 
	one expects some difference between resummed and fixed-order 
	predictions due to a strong sensitivity to unphysically low momentum scales 
	\cite{Salam:2021tbm}.

	For our theory predictions we match the \PDF{} order with the perturbative cross-section order for 
	 consistency, using \texttt{MSHT20} \PDF{} fits with $\alpha_s(m_Z)=0.118$ 
	 \cite{Bailey:2020ooq,McGowan:2022nag}.
	  This is particularly important for the logarithmic accuracy in the 
	 resummation.
         Uncertainties associated with missing higher order effects are estimated by performing
	 scale variations following the procedure of ref.~\cite{Neumann:2022lft} for the
	 Drell-Yan process.   These are symmetrized based on the maximum excursion for simplicity.
	 We also take into account uncertainties from the \PDF{} determination, which for the case of the \texttt{MSHT20} approximate \NNNLO{} \PDF{} 
         set \cite{McGowan:2022nag} accounts for missing 
	 higher-order effects within the \PDF{} in addition to uncertainty arising from the fitting procedure.  
	 Furthermore, as discussed in \cref{sec:cutoffeffects},
         our $\alpha_s^3$ results have an additional uncertainty of $0.5\%$ that covers our remaining
	 uncertainty  due to cutoff effects in the slicing procedure and matching corrections. 
	 
	\begin{figure*}
		\centering
		\includegraphics[]{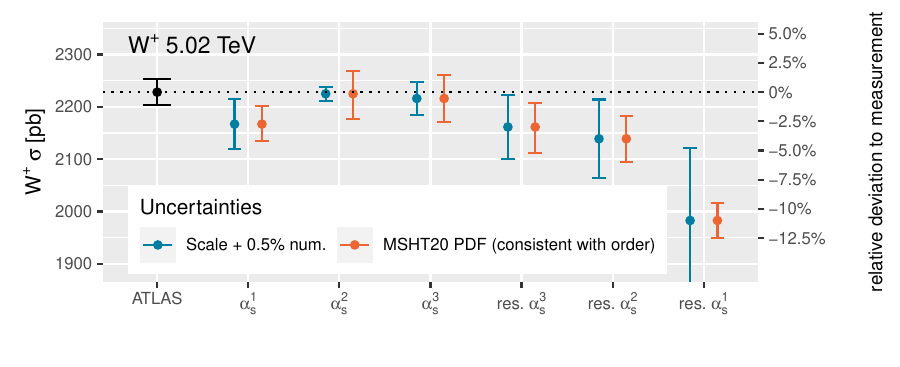}
		\caption{$W^+$ cross-sections in comparison with the \ATLAS{} \SI{5.02}{\TeV} measurement
			\cite{ATLAS-CONF-2023-028}. Error bars show uncertainties from scale variation and from 
			the \texttt{MSHT20} \PDF{} sets corresponding to the perturbative order. The 
			$\alpha_s^3$ results have an additional numerical and cutoff uncertainty of $0.5\%$ 
			that we added linearly to the scale uncertainties for display.}
		\label{fig:fiducial5TeV_msht}
	\end{figure*}
	
        Our results, and comparison to the ATLAS measurement, are shown in
	\cref{fig:fiducial5TeV_msht}.\footnote{We have added the ATLAS systematic and statistical experimental 
	uncertainties in quadrature.}		
	We find that perturbative corrections are small for both the fixed-order and 
	resummed results. Ultimately this is due to the effect of the (approximate) \NNNLO{} \PDF{}s, 
	which we discuss in the following. Scale 
	uncertainties at \NNNLO{} (<1\%) are at the level of \NNLO{} uncertainties, which is a feature 
	already observed in the literature in both inclusive and differential cases 
	\cite{Duhr:2020sdp,Chen:2022lwc}.
	On the other hand,
	scale uncertainties for our resummed results consistently decrease, but are still about $2\%$ 
	at order $\alpha_s^3$. 
	
        The $\alpha_s^3$ fixed-order and resummed cross-sections are marginally 
	compatible, and overall uncertainties from both predictions are still too large to indicate a 
	significant difference that would indicate the need for resummation 
	\cite{Salam:2021tbm}. In addition, the experimental precision is limited by about a 1\% luminosity 
	uncertainty and is compatibly with both predictions.
	Of course a direct window on this issue comes from a comparison of predictions
	with the measurement of the $q_T^W$ distribution since the bulk of the cross-section 
        comes from the region of small $q_T$ where resummation is required.  We defer this discussion
	to \cref{sec:ptW}.
	
	\paragraph{\PDF{} dependence.}
	We now extend our discussion by considering \PDF{}s from different groups, where it is not possible
	to consistently match the order in $\alpha_s$ of the \PDF{} fit to that of the perturbative calculation.
	In this section we therefore use the same \PDF{} set at each order of the perturbative 
	calculation.
	We will compare results using the \texttt{MSHT20nnlo\_as118} \PDF{} set with those from the
	\NNLO{} determinations \texttt{MSHT20nnlo\_as118}, \texttt{CT18NNLO} \cite{Hou:2019qau} and 
	\texttt{NNPDF40nnlo\_as\_0118} \cite{NNPDF:2021njg}.
	
	\begin{figure*}
		\centering
		\includegraphics[]{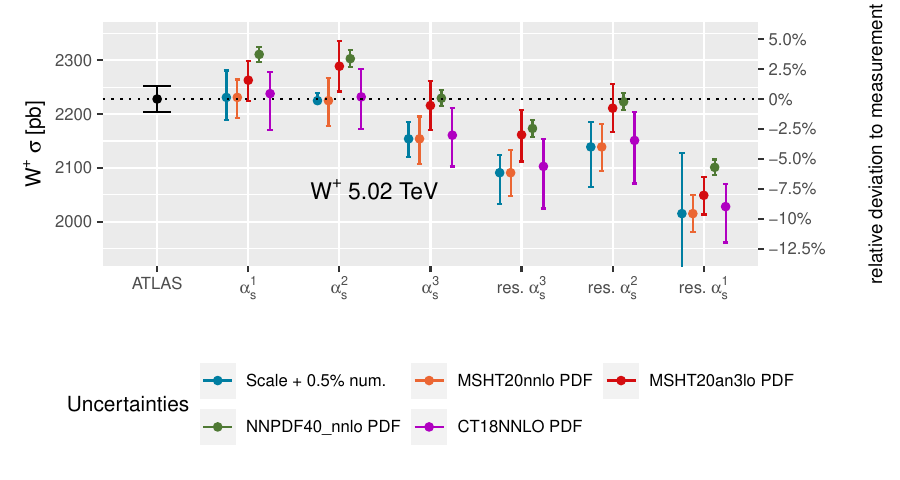}
		\caption{$W^+$ cross-sections in comparison with the \ATLAS{} \SI{5.02}{\TeV} measurement
			\cite{ATLAS-CONF-2023-028}. Error bars show uncertainties from scale variation and from 
			different \PDF{} sets with their respective uncertainties. The 
			$\alpha_s^3$ results have an additional numerical and cutoff uncertainty of $0.5\%$ 
			that we have added linearly to the scale uncertainties for display. }
		\label{fig:fiducial5TeV}
	\end{figure*}
	
        Our results are shown in \cref{fig:fiducial5TeV}.
	We first focus on the fixed-order and resummed $\alpha_s^3$ predictions using 
	\texttt{MSHT20} at \NNLO{} and a\NNNLO{}. Since the same data was used in both 
	\PDF{} fits, the difference between these predictions therefore solely results from 
	higher-order corrections in the 
	\PDF{} 	and the inclusion of \NNNLO{} K-factors in the predictions of some cross-sections. The 
	a\NNNLO{} \PDF{} increases the $\alpha_s^3$ results by about 3\% compared to using
	the \NNLO{} \PDF{}. This is a significant 
	deviation also in terms of the \PDF{} uncertainties. Without taking into account the a\NNNLO{} 
	\PDF{}s we would conclude a cross-section decrease of about 3.2\% at fixed-order and about 
	2.3\% using the resummed result.\footnote{Note that our numerical and cutoff uncertainty is about 
	0.5\%.}  This is similar in size to the effects observed in previous calculations of this process
	more inclusively, where the same \PDF{} set has been employed across all orders of the
	calculation \cite{Duhr:2020sdp,Chen:2022lwc}. 
	
	The size of the a\NNNLO{} \PDF{} effects, and the delicate cancellation between different 
	partonic channels,
	indicates that a consistent order is important, see also \cref{fig:fiducial5TeV_msht}. On the 
	other hand the \texttt{NNPDF4.0} \NNLO{} \PDF{} set is consistent with the larger 
	cross-sections of the \texttt{MSHT20} a\NNNLO{} \PDF{}. This leads to the question of what impact 
	\NNNLO{} effects will have in a \texttt{NNPDF4.0} fit. Judging by the pattern in 
	\texttt{MSHT20} 
	one would expect a sizeable positive shift, which would lead the fixed-order prediction to  
	overshoot the measurement. From this it is clear that the 
	improvement of \PDF{}s is a priority for precision predictions and measurements of this process.
	
	To conclude the discussion of the fiducial cross-section, we find that theory uncertainties are 
	overall at the level of $3$-$5$\% for the $\alpha_s^3$ $W$-boson cross-section. This 
	includes 
	scale uncertainties, the envelope of \PDF{} uncertainties and the difference between 
	fixed-order and resummed results. All uncertainties are at the same level and it will require 
	effort in 
	all directions to significantly reduce the overall theoretical uncertainty. In particular it will 
	require careful investigation of \PDF{}s in terms of higher-order effects and systematics 
	as well as the estimation of statistical uncertainties, which vary significantly between modern 
	\PDF{} sets. The comparison of differential cross-sections is likely to shed further light on 
	these issues, which we discuss further when comparing differential \PDF{} uncertainties in 
	\cref{sec:diffpdf}.
	
	\subsection{Transverse momentum distributions}
	\label{sec:ptW}

	Moving towards differential quantities, we show the $W^+$ transverse momentum distribution at 
	$\sqrt{s}=\SI{5.02}{\TeV}$ in \cref{fig:pt34}. Results for $W^-$ production can be found in 
	\cref{fig:pt34_Wm} in the appendix. To highlight the effect of short-distance corrections, 
	we use the {\texttt{MSHT20nnlo\_as118} \PDF{} set for all orders of our predictions in this and 
	in \cref{sec:mtptl}.
	
	\begin{figure}
		\centering
		\includegraphics{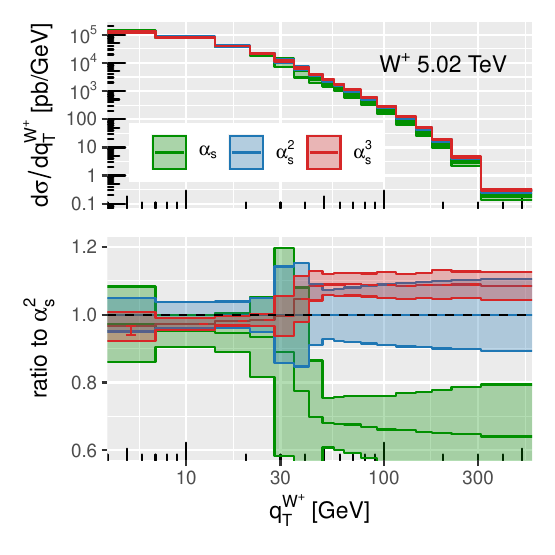}
		\caption{$W^+$ transverse momentum distribution at \SI{5.02}{\TeV} using the \PDF{} set 
		\texttt{MSHT20nnlo\_as118} throughout.}
		\label{fig:pt34}
	\end{figure}
	
	We neglect negative matching corrections of up to about 3\% below \SI{3.16}{\GeV} from $\alpha_s^3$ 
	contributions. This can amount to an 
	effect of up to about 1.5\% in the first bin in the experimental distribution which ranges up 
	to \SI{7}{\GeV}. We show the effect of neglected matching corrections with an additional 
	error bar. Since it is expected that the effect of the neglected matching corrections is 
	negative we 
	plot an error bar of 1.5\% only in the negative direction, the amount by which the 
	predictions could shift downwards. Note that our numerical precision in the first bin is 
	similar, at the level of 1\%. 
	
	Furthermore, towards very small $q_T$ scale uncertainty estimates become potentially unreliable, 
	since a downward variation requires a cutoff to not reach into the non-perturbative regime. 
	This further motivates a symmetrization of uncertainty bars, in addition to the fact that 
	distinguishing between and up- and downwards scale variation is unphysical.
	
	Overall the relative corrections are, as expected, very similar to our Drell-Yan results 
	\cite{Neumann:2022lft} with about 10\% corrections in the tail from fixed-order, and smaller 
	corrections at small $q_T$ through higher-order resummation. Uncertainties consistently 
	decrease at higher orders, including the uncertainty associated with  
	transitioning between fixed-order and resummation around \SIrange{30}{50}{\GeV} which shrinks
	considerably at higher orders. The uncertainty bands almost completely overlap between 
	$\alpha_s^3$ and $\alpha_s^2$, indicating a stabilization of the perturbative series over the 
	whole range. \NNNLO{} effects in the \PDF{} mostly cause a positive shift at small $q_T$ in the 
	first two bins, see \cref{sec:diffpdf}.
	
	In \cref{fig:WpWmratio} we show the ratio of normalized $W^+$ to $W^-$ transverse momentum 
	distributions. For this distribution we were able to digitize the plotted \ATLAS{} 
	measurements in 
	ref.~\cite{ATLAS-CONF-2023-028}. We find that the predictions at all orders are compatible within $1\%$ 
	of numerical noise. We also find excellent agreement with the measurement, as already observed 
	in 
	ref.~\cite{ATLAS-CONF-2023-028}.
	
	\begin{figure}
	\centering
	\includegraphics{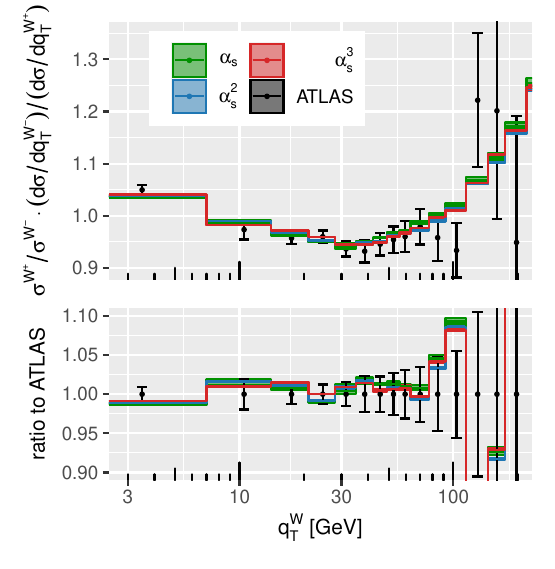}
	\caption{Ratio of $W^+/W^-$ transverse momentum distributions in comparison with the 
	\SI{5.02}{\TeV} \ATLAS{} measurement \cite{ATLAS-CONF-2023-028} using the \PDF{} set 
	\texttt{MSHT20nnlo\_as118} throughout.}
	\label{fig:WpWmratio}
	\end{figure}	
	
	The agreement of the three perturbative orders within numerical uncertainties indicates that to 
	estimate uncertainties one should compute the ratios in a correlated way, as we have done. 
	This leads to 
	uncertainties smaller than 1\%, negligible in comparison with measurement uncertainties and 
	their difference to the predictions.

	\subsection{Transverse mass and charged lepton transverse momentum distributions}
	\label{sec:mtptl}
	 
	We present the transverse mass distribution for $W^+$ in \cref{fig:mt34} and for $W^-$ in 
	\cref{fig:mt34_Wm} in the appendix. While this 
	distribution could comfortably be computed at fixed order, we only show predictions including 
	the effect of $q_T$ resummation here. 
	
	Perturbative corrections at $\alpha_s^3$ are flat in the 
	peak region and therefore, as expected from the total fiducial results presented earlier, 
	negative and about $2\%$. Note that this statement is based on using \NNLO{} \PDF{}s throughout. 
	As we will show in \cref{sec:diffpdf}, moving towards the a\NNNLO{} set, one observes a 
	considerable shift of about 5\% below the peak.
	
	\begin{figure}
		\centering
		\includegraphics{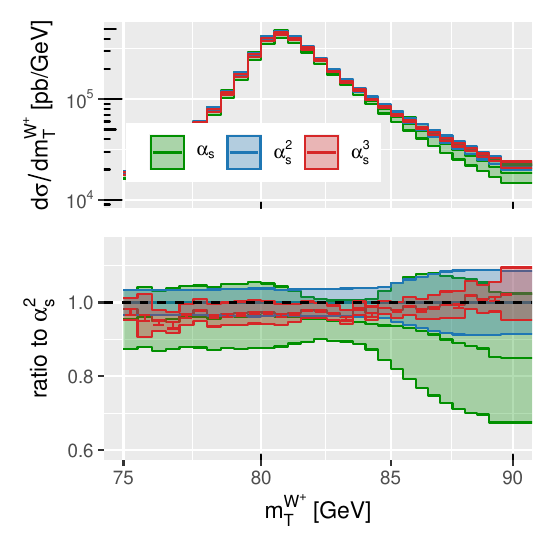}
		\caption{$W^+$ transverse mass distribution  using the \PDF{} set 
			\texttt{MSHT20nnlo\_as118} throughout.}
		\label{fig:mt34}
	\end{figure}
		
	In \cref{fig:ptl} we show the $W^+$ lepton transverse momentum distribution at different orders 
	in $\alpha_s$ (c.f. \cref{fig:ptl_Wm} in the appendix for $W^-$).
	This distribution is particularly important for the 
	$W$-boson mass determination, since it is sensitive to $m_W$ and does not depend on 
	missing-energy estimations. While 
	typically used in template fits, a new asymmetry observable based on this distribution
	has recently been proposed in ref.~\cite{Rottoli:2023xdc}.
	
	Within numerical bin-to-bin fluctuations the $\alpha_s^3$ corrections are flat 
	with smaller uncertainties. This statement also holds while using the a\NNNLO{} \PDF{}s and 
	other sets, at least below the peak region.
	
	\begin{figure}
		\centering
		\includegraphics{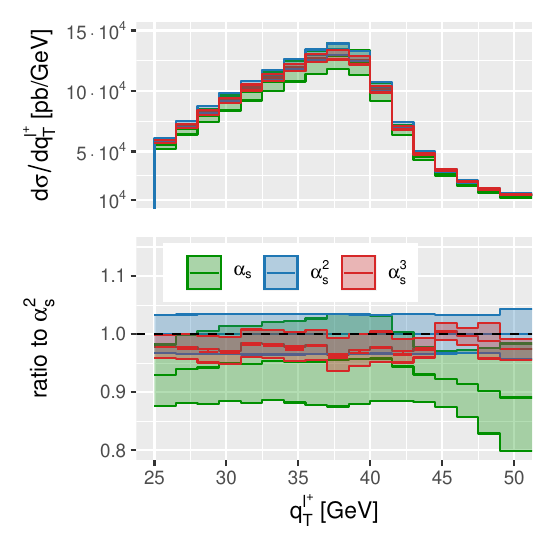}
		\caption{$W^+$ lepton transverse momentum distribution  using the \PDF{} set 
			\texttt{MSHT20nnlo\_as118} throughout.}
		\label{fig:ptl}
	\end{figure}
	
	\subsection{Impact of \PDF{}s}
	\label{sec:diffpdf}
	
	As already indicated by the fiducial cross-sections in \cref{sec:fidcross}, \PDF{}s are 
	among the biggest limitation in precise predictions. Most insight will be gained by studying 
	differential distributions.
	
	In \cref{fig:pdferr} we show the impact of four modern \PDF{} sets for the $W^+$ transverse 
	momentum, charged lepton transverse momentum, and transverse mass distributions. These relative 
	\PDF{} uncertainties are computed using $\alpha_s^2$ matrix elements. The differences to 
	$\alpha_s^3$ are at the per-mille level and insignificant for this discussion. Even using 
	$\alpha_s$ matrix elements leads to qualitatively the same conclusions \cite{Campbell:2019dru}.
	Further, the results for $W^-$ production are virtually the same, but are included for 
	completeness in the appendix in \cref{fig:pdferr_Wm}.
	
	Effects from different \PDF{} sets can be significant, depending on the distribution and region
	up to 10\%. The \texttt{MSHT} and \texttt{CTEQ} \NNLO{} \PDF{} sets are broadly similar, which 
	are the sets considered in the \ATLAS{} study in ref.~\cite{ATLAS-CONF-2023-028}, in 
	addition to \texttt{NNPDF3.1}.
	
	The most interesting comparison is between \texttt{MSHT} \NNLO{} and \texttt{MSHT} 
	a\NNNLO{} as a higher-order effect, and then considering \texttt{NNPDF4.0} \NNLO{}.
	We find significant shape changes for several distributions when utilizing these
	\PDF{} sets.
		
	The effects of using  \texttt{MSHT} a\NNNLO{} are even more important differentially than inclusively, 
	inducing a significant cross-section increase below the peaks for the transverse mass and 
	lepton $q_T$ distributions, while dropping off beyond. 
	In the $W$-boson transverse-momentum distribution the most significant change 
	is a positive shift of about 7\% in the first bin containing the Sudakov peak. \PDF{} 
	uncertainties even range up to 10\%. Clearly such a range can be constrained within \QCD{} 
	uncertainties in future fits, and even precise Drell-Yan measurements and predictions 
	\cite{Neumann:2022lft} will constrain this significantly.
	
	Predictions using \texttt{NNPDF4.0} \NNLO{} for $m_t^W$ and $q_T^l$ are much flatter with 
	respect to \texttt{MSHT20} \NNLO{}, except for $q_T^W$, which predicts a similar enhancement of 
	about 5\% in the first bin, but drops off slower than \texttt{MSHT20} a\NNNLO{}.
	
		\begin{figure}
	\centering
	\begin{subfigure}[b]{0.45\textwidth}
		\includegraphics[width=\textwidth]{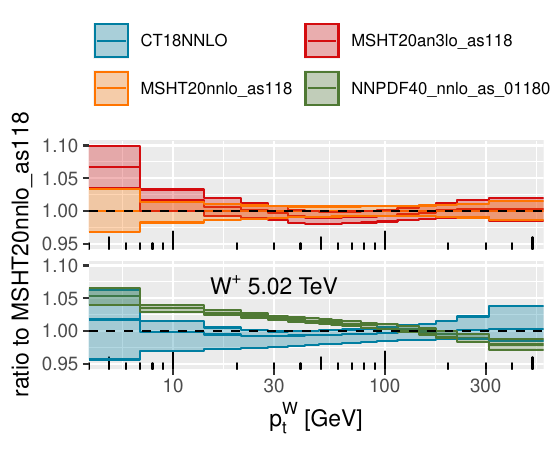}
	\end{subfigure}
	\\
	\begin{subfigure}[b]{0.45\textwidth}
		\includegraphics[width=\textwidth]{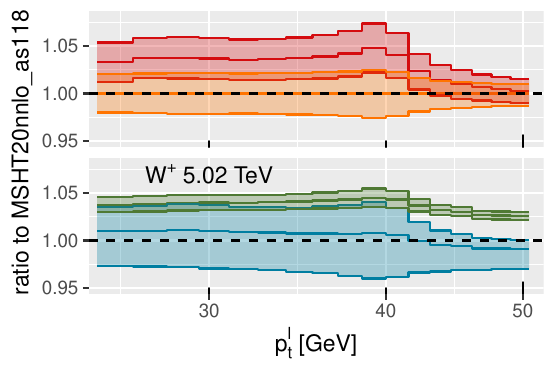}
	\end{subfigure}
	\\
	\begin{subfigure}[b]{0.45\textwidth}
		\includegraphics[width=\textwidth]{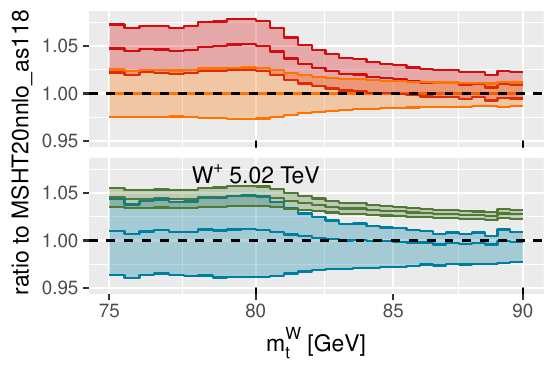}
	\end{subfigure}
	\caption{Relative \PDF{} uncertainties of different observables for $W^+$. Note that 
	\texttt{MSHT20an3lo} includes uncertainties from missing higher orders, which are not 
	included in the other sets.}
	\label{fig:pdferr}
	\end{figure}
	
	\section{Conclusions \& Outlook}
	\label{sec:conclusion}
	
	In recent years the experimental precision of $Z$ and $W$-boson production has reached new 
	levels at the \LHC{}. In particular this has been achieved through better measurements of the 
	luminosity uncertainty which is now down to $1\%$. Precise measurements of $W$-boson kinematics 
	enter many Standard Model inputs like the weak mixing angle, parton distribution functions, and 
	$W$-boson mass.  At the same time, theoretical calculations have become more advanced,
	reaching new levels of precision in fixed-order and resummed predictions, and allowing
	for more refined \PDF{} determinations.
	However these calculations have presented new challenges for performing precision
	measurements.
	\NNNLO{} \QCD{} corrections are surprisingly large, at the level of minus $2$-$3\%$ 
	\cite{Duhr:2020sdp} (disregarding the effect of \NNNLO{} \PDF{}s), and more statistically 
	precise 
	\PDF{} fits begin to reveal systematic discrepancies that are challenging to reconcile.
	
	\paragraph{Conclusions.}
	
	In this paper we have presented a calculation of $W$-boson production at the level of $\alpha_s^3$ at fixed-order 
	and including the effect of $q_T$ resummation. We find that the size of the corrections
	depends crucially on whether higher-order corrections are included consistently in the
	short-distance calculation and the \PDF{}s. We considered \texttt{MSHT20} 
	\PDF{}s, which, taken at \NNLO{}, lead to results with about $-3\%$ corrections at 
	$\alpha_s^3$. On the other hand using the a\NNNLO{} version \cite{McGowan:2022nag} to 
	consistently match the 
	matrix-element we find that corrections are less than half a percent, which is about our 
	numerical precision. When comparing with data, it seems important to consistently match orders.
	Other \PDF{} sets like {\abbrev NNPDF4.0} overshoot the experimental measurement at \NNLO{}, 
	but then match it at \NNNLO{} due to the large corrections. Whether \NNNLO{} effects in other 
	\PDF{} fits lead to a similar increase as for \texttt{MSHT20}, and how the systematic 
	differences between them are resolved, will have to be studied in detail in the future.
	
	Our fiducial cross-sections with $q_T$ resummation are overall smaller than at fixed order, but 
	we find larger missing-higher-order uncertainties of about $2.5\%$. Here in particular the 
	higher-order {\abbrev DGLAP} evolution is important to achieve full \NFLL{} accuracy and it is 
	therefore important to match the \PDF{} order.
		
	We calculated differential distributions for the $W^\pm$ transverse momentum, charged lepton 
	transverse momentum and transverse mass distributions. While data for such distributions has 
	not been made public yet, we have digitized the $W^+/W^-$ transverse momentum ratio and found 
	agreement, confirming earlier results at lower orders.
	
	We have illustrated the impact of different \PDF{} sets and their uncertainties for these three 
	kinematic distributions. We find large shape differences between \texttt{MSHT20} \NNLO{} and 
	a\NNNLO{}, but also between \texttt{NNPDF40\_nnlo} and the other sets considered. The 
	differences and shape changes reach  
	5--10\%. It is therefore clear that the precise measurements will further strongly constrain 
	\PDF{}s.
	
	In the future it will be interesting to include the effect of non-perturbative corrections in 
	the context of transverse-momentum-dependent \PDF{}s 
	\cite{Moos:2023yfa,Scimemi:2019cmh,Bacchetta:2019sam}. In the typical formalisms the 
	disentangling of perturbative and non-perturbative effects in a model-independent way is 
	difficult \cite{Ebert:2022cku}, but which is simpler here since no direct Landau pole 
	regularization is needed \cite{Becher:2013iya}. A further phenomenological avenue will be to 
	study the effect of higher-order resummation on angular coefficients which are used in 
	experimental studies to dress parton-shower results with higher-order corrections.
		
	Our calculation including the matching to $W$+jet production at \NNLO{} will be 
	publicly available the upcoming \CuTeMCFM{} release and allows for theory-data comparison at 
	the state-of-the-art level.

	\paragraph{Calculational outlook.}
	
	While our calculation at \NNNLO{} is computationally expensive compared to \NLO{} or \NNLO{} 
	results, which can be computed in a short time on modern multi-core desktops, it is relatively 
	low compared to other state-of-the-art calculations \cite{FebresCordero:2022psq}. The precision 
	reached in this paper is about $0.5\%$ for total fiducial cross-sections, and is limited by the 
	double-real $W$+jet \NNLO{} calculation at small $q_T$. We have used about 1500 NERSC 
	Perlmutter node hours for this, which translates into about 100,000 core hours.
	
	Unfortunately we are roughly limited to a precision of $0.5\%$ for the chosen set of fiducial 
	cuts which have large matching corrections in the case of resummation, and require small 
	$\qtcut{}$ values to reach $q_T$-subtraction asymptotics at fixed order.
 	In our nested slicing approach such small 
	$\qtcut{}$ values come at the price of correspondingly smaller $\taucut{}$ in the $W$+jet 
	\NNLO{} slicing 
	calculation.
	 For example, we estimate that decreasing the cutoff to less than \SI{1}{\GeV} will require an 
	 order of magnitude smaller $\taucut$. Since Monte-Carlo integration uncertainties decrease  
	 asymptotically 
	 only like 
	$1/\sqrt{N}$, where $N$ is the number of calls (or runtime), one quickly reaches the limit of 
	reasonable runtimes. Future improvements based on local subtractions for the $W$+jet \NNLO{} 
	calculation, and ultimately for the \NNNLO{} cross-section will naturally improve this.
	
	In practice the numerical uncertainty of $0.5\%$ does not pose a problem. In the $W$-boson 
	$q_T$ distribution it only affects the first bin, typically at the level of $1\%$ but
	dependent on the exact extent of the bin.
	In our 
	other resummed distributions it is smeared out and still small compared to our estimation of 
	missing-higher-order uncertainties.

	\paragraph{Acknowledgments.} We 
would like to thank {\abbrev NERSC} for use of the Perlmutter 
supercomputer that enabled this calculation.
This manuscript has been authored by Fermi Research Alliance, LLC under Contract No.
DE-AC02-07CH11359 with the U.S. Department of Energy, Office of Science, Office of High Energy 
Physics.
Tobias Neumann is supported by 
the United States Department of Energy under Grant Contract DE-SC0012704.
This research used resources of the National Energy Research Scientific Computing Center (NERSC), a 
U.S. Department of Energy Office of Science User Facility located at Lawrence Berkeley National 
Laboratory, operated under Contract No. DE-AC02-05CH11231 using NERSC award HEP-ERCAP0023824 
\enquote{Higher-order calculations for precision collider phenomenology}.

	\appendix
	
	\section{Results for $W^-$ production}
	\label{app:additional}
	
	A comparison of the fiducial cross-section 
        measurement at $\sqrt{s}=\SI{5.02}{\TeV}$ with our theoretical predictions for $W^-$ production
	is shown in \cref{fig:fiducial5TeV_Wm}.
	The relative perturbative corrections to the $W^-$-boson transverse momentum 
	(\cref{fig:pt34_Wm}), the $W^-$-boson transverse mass 
	(\cref{fig:mt34_Wm}), and the charged lepton 
	transverse momentum (\cref{fig:ptl_Wm}) 
	distributions are very similar to the corresponding $W^+$ results in the main text, see 
	\cref{fig:pt34,fig:ptl,fig:mt34}. The impact of different \PDF{} fits with uncertainties for 
	$W^-$ distributions is 
	shown in \cref{fig:pdferr_Wm}. We find that the relative positions of the individual 
	predictions shift only slightly compared to $W^+$ production.
	
	\begin{figure*}
	\centering
	\includegraphics[]{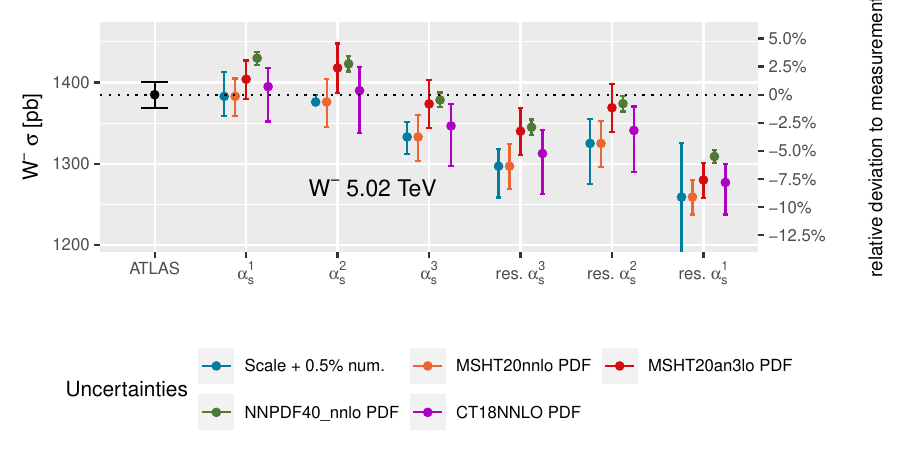}
	\caption{$W^-$ cross-sections in comparison with the \ATLAS{} \SI{5.02}{\TeV} measurement
		\cite{ATLAS-CONF-2023-028}. Error bars show uncertainties from scale variation and from 
		different \PDF{} sets with their respective uncertainties. The 
		$\alpha_s^3$ results have an additional numerical and cutoff uncertainty of $0.5\%$ that we 
		added linearly to the scale uncertainties. }
	\label{fig:fiducial5TeV_Wm}
	\end{figure*}

	\begin{figure}
		\centering
		\includegraphics{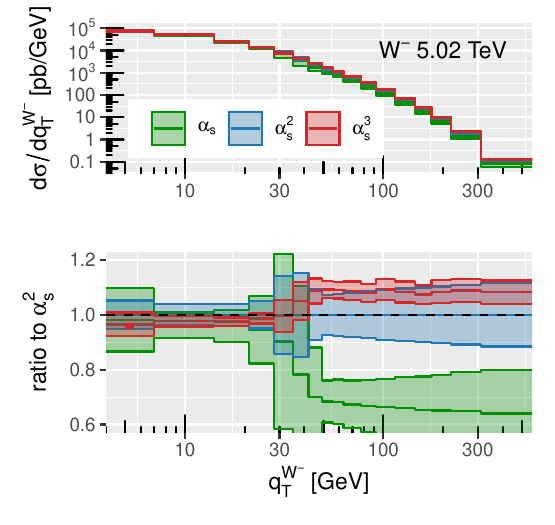}
		\caption{$W^-$ transverse momentum distribution at \SI{5.02}{\TeV}.}
		\label{fig:pt34_Wm}
	\end{figure}
	
		\begin{figure}
		\centering
		\includegraphics{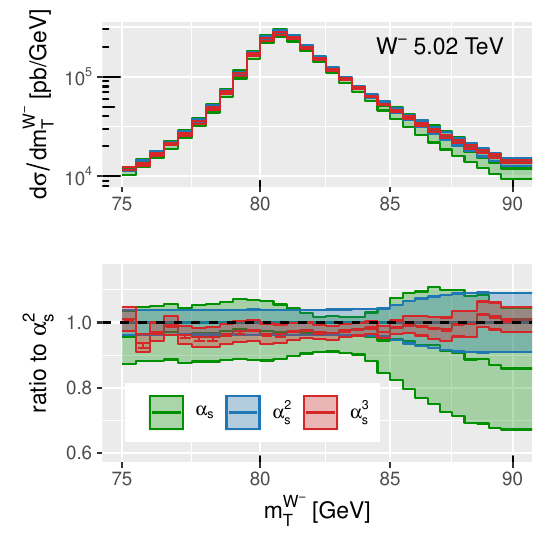}
		\caption{$W^-$ transverse mass distribution at \SI{5.02}{\TeV}.}
		\label{fig:mt34_Wm}
	\end{figure}
	
	\begin{figure}
		\centering
		\includegraphics{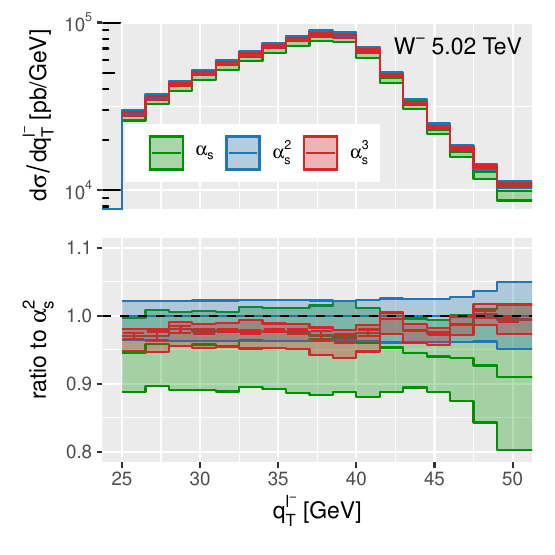}
		\caption{$W^-$ electron transverse momentum distribution at \SI{5.02}{\TeV}.}
		\label{fig:ptl_Wm}
	\end{figure}
	
		\begin{figure}
	\centering
	\begin{subfigure}[b]{0.45\textwidth}
		\includegraphics[width=\textwidth]{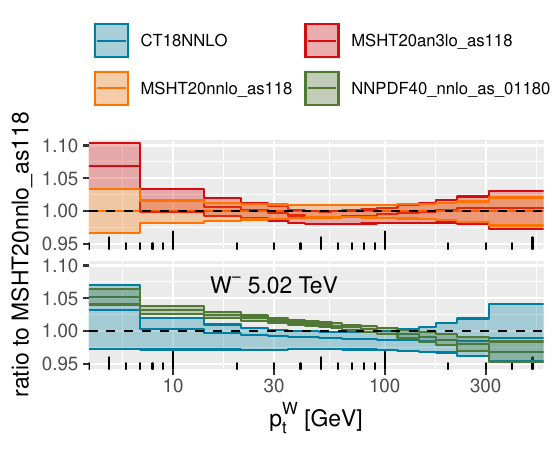}
	\end{subfigure}
	\\
	\begin{subfigure}[b]{0.45\textwidth}
		\includegraphics[width=\textwidth]{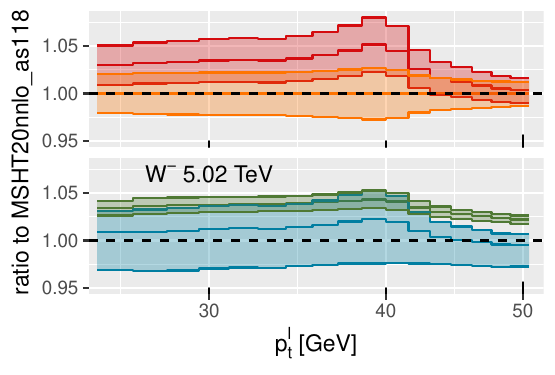}
	\end{subfigure}
	\\
	\begin{subfigure}[b]{0.45\textwidth}
		\includegraphics[width=\textwidth]{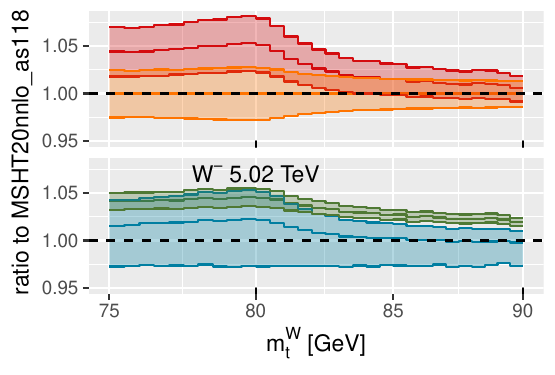}
	\end{subfigure}
	\caption{Relative \PDF{} uncertainties of different observables for $W^-$. Note that 
		\texttt{MSHT20an3lo} includes uncertainties from missing higher orders, which are 
		not included in the other sets.}
	\label{fig:pdferr_Wm}
\end{figure}	

	\cleardoublepage
	
	\bibliographystyle{JHEP}
	\bibliography{refs}
	
\end{document}